\pdfoutput=1

\documentclass[11pt,twoside,a4paper,cmspaper,final,collab]{cms-tdr}
\def\svnVersion{80402}\def\cmsCernNo{2011-092}\def\cmsCernDate{\today}\def\cmsMessage{Submitted to the Journal of High Energy Physics}

\begin{document}\cmsNoteHeader{HIN-10-001}

\hyphenation{had-ron-i-za-tion}
\hyphenation{cal-or-i-me-ter}
\hyphenation{de-vices}
\RCS$Revision: 71567 $
\RCS$HeadURL: svn+ssh://alverson@svn.cern.ch/reps/tdr2/papers/HIN-10-001/trunk/HIN-10-001.tex $
\RCS$Id: HIN-10-001.tex 71567 2011-07-24 15:08:35Z alverson $
\newcommand {\ave}[1]   {\ensuremath{\left< #1 \right>}}
\newcommand {\abs}[1]   {\ensuremath{\left| #1 \right|}}
\newcommand {\prn}[1]   {\ensuremath{\left( #1 \right)}}
\newcommand {\sqb}[1]   {\ensuremath{\left[ #1 \right]}}
\newcommand {\lsim}     {\,{\buildrel < \over {_\sim}}\,}
\newcommand {\gsim}     {\,{\buildrel > \over {_\sim}}\,}
\newcommand {\cov}      {\ensuremath{\mathrm{cov}}}

\newcommand {\hrefurl}[1]{\href{#1}{\mbox{#1}}}
\newcommand {\fig}[1]{Fig.~\ref{#1}}
\newcommand {\figs}[2]{Fig.~\ref{#1} and~\ref{#2}}
\newcommand {\Fig}[1]{Figure~\ref{#1}}
\newcommand {\eq}[1]{Eq.~\ref{#1}}
\newcommand {\eqs}[2]{Eq.~\ref{#1} and~\ref{#2}}
\newcommand {\Eq}[1]{Equation~\ref{#1}}
\newcommand {\Eqs}[2]{Equations~\ref{#1} and~\ref{#2}}
\newcommand {\sect}[1]{Sect.~\ref{#1}}
\newcommand {\sects}[2]{Sect.~\ref{#1} and~\ref{#2}}
\newcommand {\Sect}[1]{Section~\ref{#1}}
\newcommand {\Sects}[2]{Sections~\ref{#1} and~\ref{#2}}
\newcommand {\chap}[1]{Ch.~\ref{#1}}
\newcommand {\chaps}[2]{Ch.~\ref{#1} and~\ref{#2}}
\newcommand {\Chap}[1]{Chapter~\ref{#1}}
\newcommand {\tab}[1]{Table~\ref{#1}}
\newcommand {\tabs}[2]{Tables~\ref{#1} and~\ref{#2}}
\newcommand {\Tab}[1]{Table~\ref{#1}}
\newcommand {\peq}[1]{Eq.~\ref{#1} on page~\pageref{#1}}
\newcommand {\pEq}[1]{Equation~\ref{#1} on page~\pageref{#1}}
\newcommand {\psect}[1]{Sect.~\ref{#1} on page~\pageref{#1}}
\newcommand {\pSect}[1]{Section~\ref{#1} on page~\pageref{#1}}
\newcommand {\pfig}[1]{Fig.~\ref{#1} on page~\pageref{#1}}
\newcommand {\pFig}[1]{Figure~\ref{#1} on page~\pageref{#1}}
\newcommand {\ptab}[1]{Table~\ref{#1} on page~\pageref{#1}}
\newcommand {\pTab}[1]{Table~\ref{#1} on page~\pageref{#1}}
\newcommand {\page}[1]{page~\pageref{#1}}
\newcommand {\appndx}[1]{Appendix~\ref{#1}}

\newcommand {\roots}    {\ensuremath{\sqrt{s}}}
\newcommand {\dndy}     {\ensuremath{dN/dy}}
\newcommand {\dnchdy}   {\ensuremath{dN_{\mathrm{ch}}/dy}}
\newcommand {\dndeta}   {\ensuremath{dN/d\eta}}
\newcommand {\dnchdeta} {\ensuremath{dN_{\mathrm{ch}}/d\eta}}
\newcommand {\dndpt}    {\ensuremath{dN/d\pt}}
\newcommand {\dnchdpt}  {\ensuremath{dN_{\mathrm{ch}}/d\pt}}
\newcommand {\deta}     {\ensuremath{\Delta\eta}}
\newcommand {\dphi}     {\ensuremath{\Delta\phi}}

\newcommand {\pp}    {\mbox{pp}}
\newcommand {\ppbar} {\mbox{p\={p}}}
\newcommand {\pbarp} {\mbox{p\={p}}}
\newcommand {\PbPb}  {\mbox{Pb-Pb }}

\newcommand{\m}{\ensuremath{\,\text{m}}\xspace}

\newcommand {\naive}    {na\"{\i}ve}
\providecommand{\GEANT} {\textsc{geant}\xspace}

\def\d{\mathrm{d}}

\cmsNoteHeader{AN-10-365} 
\title{Dependence on pseudorapidity and on centrality of charged hadron production in PbPb collisions at $\sqrt{s_{_{\rm NN}}}$~=~2.76~TeV}

\author[cern]{Yen-Jie Lee, Krisztian Krajczar, Gunther Roland, Gabor Veres, Edward Wenger, Yetkin Yilmaz
}

\date{\today}

\abstract{
A measurement is presented of the charged hadron multiplicity in
hadronic PbPb collisions, as a function of pseudorapidity and
centrality, at a collision energy of 2.76~TeV per nucleon pair. The
data sample is collected using the CMS detector and a
minimum-bias trigger, with the CMS solenoid off. The number of
charged hadrons is measured both by counting the number of
reconstructed particle hits and by forming hit doublets of pairs of
layers in the pixel detector. The two methods give consistent results.
The charged hadron multiplicity density, $dN_{\rm ch}/d\eta|_{\eta=0}$,
for head-on collisions is found to be $1612\pm 55$,
where the uncertainty is dominated by systematic effects.
Comparisons of these results to previous measurements and to various
models are also presented.}

\hypersetup{%
pdfauthor={CMS Collaboration},%
pdftitle={Dependence on pseudorapidity and centrality of charged hadron production in PbPb collisions at a nucleon-nucleon centre-of-mass energy of 2.76 TeV},%
pdfsubject={CMS},%
pdfkeywords={CMS, physics}}

\maketitle 

\setcounter{tocdepth}{2} 

\section{Introduction}
\label{chap:intro}

Quantum chromodynamics (QCD), the theory of strong interactions,
predicts a phase transition at high temperature between hadronic and
deconfined matter \cite{Karsch:2003jg}. Strongly interacting matter
under extreme conditions can be studied experimentally using
ultrarelativistic collisions of heavy nuclei. The field entered a new
era in November 2010 when the Large Hadron Collider (LHC) produced the
first PbPb collisions at a centre-of-mass energy per nucleon pair of
2.76~TeV. This represents an increase of more than one order of
magnitude over the highest-energy nuclear collisions previously achieved
in the laboratory. The multiplicity of charged particles produced in the
central-rapidity region is a key observable characterising the
properties of the quark-gluon matter created in these collisions
\cite{Kharzeev:2000ph}.

Nuclei are extended objects, and their collisions occur at various
impact parameters, referred to as ``centralities''. The studies of the
dependence of the charged particle density on the type of colliding
nuclei, on the centre-of-mass energy, and on the collision geometry are
important for understanding the relative contributions of hard
scattering and soft processes to particle production and provide insight
into the partonic structure of the nuclei.

In this paper we report measurements of the multiplicity density
$dN_{\rm ch}/d\eta$ of primary charged hadrons. The analysis is based on
the 2.76~TeV-per-nucleon PbPb collision data recorded by the Compact
Muon Solenoid (CMS) detector in December 2010, in runs without magnetic
field. The pseudorapidity is defined as $\eta = -\ln[\tan(\theta/2)]$
with $\theta$ the polar angle with respect to the counterclockwise beam
direction (the $z$ axis). The number of primary charged hadrons $N_{\rm
ch}$ is defined as all charged hadrons produced in an event including
decay products of particles with proper lifetimes less than 1 cm.

A detailed description of the CMS experiment can be found in 
Ref.~\cite{JINST}. The pixel tracker used for the analysis covers the 
region $|\eta|<2.5$ and a full 2$\pi$ in azimuth, with 66M detector 
channels out of which 97.5\% were functional during data taking. It 
consists of a three-layer barrel pixel detector (BPIX) and two endcap 
disks at each barrel end. Only the barrel section was used in this 
analysis. The first BPIX layer is located at a radius between 3.6 and 
5.2~cm from the beam line, the second between 6.6 and 8.0~cm, and the 
third between 9.4 and 10.8~cm. The detectors used for event selection 
are the hadron forward (HF) calorimeters, which cover the region 
$2.9<|\eta|<5.2$, the beam scintillator counters (BSC), in the range 
$3.23 <|\eta|<4.65$, and the beam pick-up timing (BPTX) devices located 
at $z=\pm176$~m from the interaction point. The operation of the CMS 
detector with zero magnetic field has the benefit of an increase in the 
acceptance for charged hadrons down to $\sim$30~\MeVc\ transverse 
momentum ($p_{\rm T}$) without the drawbacks from particles with small 
$p_{\rm T}$ curling up in the magnetic field. The nonzero $p_{\rm T}$ 
threshold is due to the $0.8$~mm-thick beryllium beampipe, which is not 
penetrable for pions and protons below $p_{\rm T}\approx$30 and 
140~\MeVc, respectively. The loss of particles due to the beampipe is 
estimated to be less than 1 percent of the produced primary charged 
hadrons.

Two analysis methods were used for the measurements of $dN_{\rm
ch}/d\eta$ as a function of $\eta$ and centrality: one uses only pixel
clusters in single BPIX layers (hit-counting method), and the other uses
doublets of pixel clusters reconstructed from pairs of BPIX layers
(tracklet method).

The application of the pixel hit-counting method is a demonstration of
the excellent pixel detector response and of its low occupancy even in
this high-multiplicity environment, as well as the absence of noise and
background. This method is not sensitive to detector misalignment or
vertex-position resolution. The tracklet method is essentially a
coincidence version of hit-counting. Using the angular coincidence
of two hits from the same particle in different layers of the BPIX has
the important feature of suppressing random noise.

The paper is organized as follows. The triggering and event selection
requirements are explained in Section~\ref{chap:triggering}, followed by
the description of the determination of the reaction centrality in
Section~\ref{section:centrality}.
Sections~\ref{chap:clusterCountingMethod} and \ref{chap:trackletMethod}
introduce the hit-counting and the tracklet methods, respectively.
The systematic uncertainties are discussed in
Section~\ref{systerr_summary}, while the final results are presented in
Section~\ref{sec:Results}.

\section{Trigger and event selection}
\label{chap:triggering} 

The expected cross section for PbPb hadronic inelastic collisions at 
$\sqrt{s_{_{NN}}}=2.76$ TeV is 7.65~b, according to the chosen 
Glauber MC parameters described in Section~\ref{section:centrality}. 
Electromagnetic interactions of the colliding nuclei at large impact 
parameter (ultraperipheral collisions, UPC) can lead to the breakup of 
one or both Pb nuclei with a much higher cross section.

Minimum-bias (hadronic inelastic) collisions were selected by the 
Level-1 trigger system combining the logical OR of two clean and highly 
efficient triggers. One of them was the BSC coincidence, which requires 
at least one segment of the BSC firing on each side of the interaction 
point. The other was an HF coincidence trigger, which requires at least 
one HF tower on each side to have deposited energies that exceed the 
readout threshold. Both triggers accept noise at a low rate (less than 
1~Hz with two noncolliding beams at full intensity), and have a very 
high efficiency (approximately 99\% after the requirement of a 
reconstructed vertex). In order to suppress noncollision-related noise, 
cosmic-ray events, radioactivation, instrumental multiple triggering 
effects, and beam background, two colliding ion bunches were required to 
be present in coincidence with each one of these triggers, using 
information from the BPTX devices. The HF and BSC coincidence triggers 
were found to be largely insensitive to single-dissociation UPC, as 
discussed at the end of this section.

The collision rate was 1.0--1.85~Hz per colliding bunch pair during the 
PbPb data taking period. Therefore, with an orbit frequency of 11\,245 
Hz, the average number of collisions per bunch crossing was 
0.9--$1.6\times10^{-4}$. There were $129\times 129$ colliding bunches 
in the LHC at the time of data taking with no CMS magnetic field.

In order to reject beam-gas interactions, large-hit-multiplicity beam 
background, and UPC, several offline event selection requirements were 
imposed:

\begin{itemize} 
\item Events containing beam-halo muons and other 
particles from upstream collisions were identified and excluded from 
the analysis, by requiring the time difference between two hits from the 
BSC stations on opposite sides of the interaction point to be within 20 
ns of the mean flight time between them (73 ns).

\item The large-multiplicity beam-background events were removed by 
requiring the compatibility of the observed pixel-cluster lengths 
(defined in Section~\ref{chap:clusterCountingMethod}) with the 
hypothesis of a PbPb interaction. This filter is the same as the one 
used in Ref.~\cite{CMS_dNdeta_pp_2}.

\item An HF coincidence requirement was imposed. At least 3 HF towers 
were required on each side of the interaction point with at least 3~GeV 
total deposited energy in each tower.

\item Furthermore, the presence of a reconstructed event vertex was 
required. The analysis methods use their corresponding analysis objects 
(pixel clusters and tracklets, respectively) to reconstruct the 
interaction point. The vertex reconstruction is only done along the 
beamline; the transverse position of the vertex is taken to be that of 
the beam axis \cite{TRK-10-005}. The methods to determine the collision 
vertex are described in Sections~\ref{sec:cluster_vertexing} and 
\ref{sec:tracklet_vertexing}. \end{itemize}

The measurement of the $dN_{\rm ch}/d\eta$ distributions was performed 
using $100\,031$ events, corresponding to an integrated luminosity of 13 
mb$^{-1}$. Correction factors were determined using simulated events 
generated with the {\sc ampt} Monte Carlo (MC)~\cite{AMPT} program. This 
program combines the {\sc hijing} event generator~\cite{HIJING} with the 
{\sc zpc} parton cascade procedure~\cite{ZPC} and the {\sc art} 
relativistic transport model~\cite{ART} for the last stage of parton 
hadronization. The default tune is used, and the simulated events are 
reconstructed with the same version of software as used to process the 
collision data. This event generator produces a larger tail in the 
multiplicity distribution than that observed in data, making the entire 
observed multiplicity region completely covered in the simulation. The 
charged hadron multiplicity in the most-central collisions is $20\%$ 
higher in {\sc ampt} than in data, but since the analysis is done in 
bins of multiplicity, it is insensitive to this difference.

The event selection for hadronic collisions was fully efficient for 
(mid)central PbPb collisions. For peripheral collisions the event 
selection efficiency was determined by comparing peripheral PbPb data 
and $\sqrt{s}=2.76$ TeV pp data with the {\sc ampt} and {\sc pythia Z2} 
\cite{pythia} simulations. Based on these studies, the total event 
selection efficiency of the minimum-bias trigger for events produced in 
hadronic PbPb interactions was found to be $(99 \pm 1)$\%.

The UPC contamination in the selected event sample was estimated using 
the photo-dissocia\-tion simulations from Ref.~\cite{UPC}. The 
single-lead photo-dissociation events were found to be 100\% rejected by 
the event selection criteria outlined above, while half of the 
double-lead photo-dissociation events are found to pass the minimum-bias 
trigger. Such a UPC contamination amounts to $(1 \pm 0.5)$\% of the 
total number of events collected and populates $\approx$15\% (5\%) of 
the 95--100\% (90--95\%) most peripheral (largest-centrality) events, 
being negligible for the remaining 0--90\% fraction of the PbPb cross 
section.

\section{Centrality determination}
\label{section:centrality}

\begin{figure}[t]
\begin {center}
\includegraphics[width=.45\textwidth]{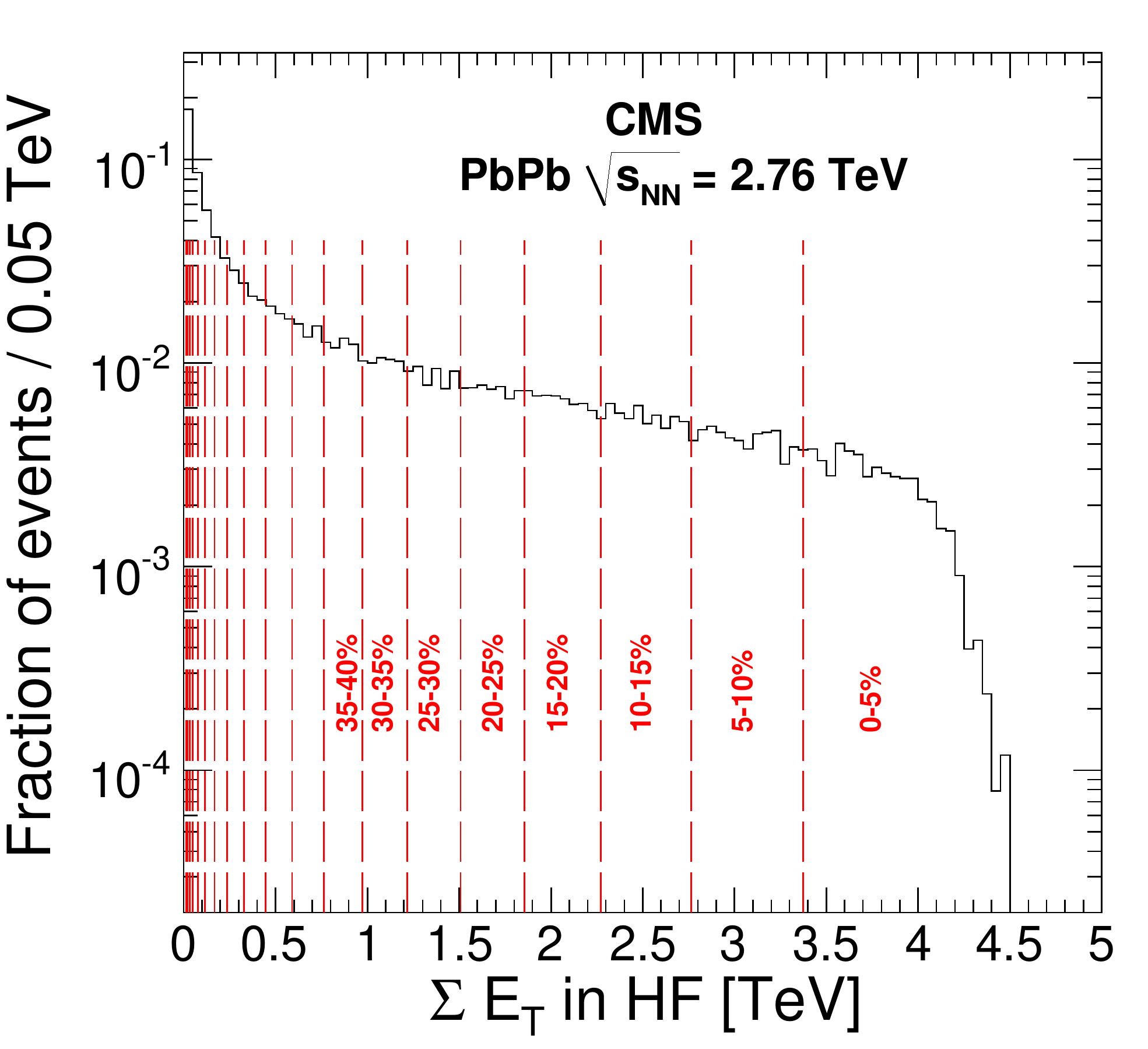}
\caption{
Distribution of the total transverse energy in the HF used to 
determine the centrality of the PbPb interactions. The centrality 
boundaries for each 5\% centrality interval are shown by the dashed 
lines.}
\label{fig:HF_cent}
\end{center}
\end{figure}

In studies with heavy ions, it is important to determine the degree of 
overlap of the two colliding nuclei, the so-called centrality of the 
interaction. Centrality is estimated using the sum of transverse energy 
in towers from both HF at positive and negative $z$ positions. The 
distribution of the total transverse energy, after the trigger 
efficiency and the UPC corrections, was used to divide the event sample 
into bins, each representing 5\% of the total nucleus-nucleus 
interaction cross section. The bin corresponding to the most central 
events (i.e.\ smallest impact parameter) is the 0--5\% bin, the next one 
is 5--10\% and so on. The distribution of the HF signal, along with the 
cuts used to define the various event classes, is shown in 
Fig.~\ref{fig:HF_cent}. The UPC are concentrated in the two 
most-peripheral bins. To avoid them completely, only the 0--90\% bins 
are used for the measurements reported in this paper.

The centrality binning using equal fractions of the total interaction 
cross section can be correlated with more detailed properties 
of the collision. The quantity of interest for this measurement is the 
total number of nucleons in the two Pb nuclei that experienced at least 
one inelastic collision, $N_{\rm part}$. The average values of $N_{\rm 
part}$ for the various centrality bins (from most-central to 
most-peripheral), together with their uncertainties, are given in 
Table~\ref{table:npart}. The $N_{\rm part}$ values were obtained using a 
Glauber MC simulation \cite{glauber,phobos_glauber} with the same 
parameters as in Ref.~\cite{HIN-10-004}. These calculations were 
translated into reconstructed centrality bins using correlation 
functions between $N_{\rm part}$ and the measured total transverse 
energy, obtained from {\sc ampt} simulated events. Different Glauber MC 
samples were produced varying the Glauber parameters within the 
uncertainties from Refs.~\cite{Vries:1987qc} and 
\cite{Glaub_syst_PHOBOS}. The variation in the final results is quoted 
as the uncertainty in $N_{\rm part}$.

\begin{table}[htb]
\centering
\caption{\label{table:npart} Average $N_{\rm part}$ values and their 
uncertainties for each PbPb centrality range defined in 5 percentile 
segments of the total inelastic cross section. The values were obtained 
using a Glauber MC simulation with the same parameters as in 
Ref.~\cite{HIN-10-004}.}
\begin{tabular}{c|cccccccccccccccccc}
\hline
\hline
Centrality & 0--5\% & 5--10\% & 10--15\% & 15--20\% & 20--25\% & 25--30\% \\ 
$N_{\rm part}$ & $381\pm2$ & $329\pm3$ & $283\pm3$ & $240\pm3$ & $203\pm3$ & $171\pm3$ \\ 
\hline
Centrality & 30--35\% & 35--40\% & 40--45\% & 45--50\% & 50--55\% & 55--60\% \\
$N_{\rm part}$ & $142\pm3$ & $117\pm3$ & $95.8\pm3.0$ & $76.8\pm2.7$ & $60.4\pm2.7$ & $46.7\pm2.3$ \\
\hline
Centrality & 60--65\% & 65--70\% & 70--75\% & 75--80\% & 80--85\% & 85--90\% \\
$N_{\rm part}$ & $35.3\pm2.0$ & $25.8\pm1.6$ & $18.5\pm1.2$ & $12.8\pm0.9$ & $8.64\pm0.56$ & $5.71\pm0.24$ \\
\hline
\hline
\end{tabular}
\end{table}

\section{Hit-counting method and corrections}
\label{chap:clusterCountingMethod}

Charged particles traversing the pixel detector deposit a certain energy 
in the silicon sensors, resulting in a proportional amount of charge 
collected in the pixel readout cells. Contiguous pixel cells with charge 
above the readout threshold are merged into a pixel cluster. A pixel 
cluster might be split into multiple clusters if one of its pixel cells 
fluctuates below the threshold. This phenomenon is called cluster 
splitting. The fraction of split clusters was estimated from the 
cluster-to-cluster distance distribution. The fractions in data and 
simulation were found to differ by less than 0.6\%. The pixel-cluster 
efficiency (i.e.\ the probability that a cluster is detected once a 
charged particle crosses a pixel-detector layer), as well as the 
fraction of large clusters split into two are important quantities for 
the measurement.

The pixel-cluster efficiency has been extensively studied in pp 
collisions \cite{CMS_dNdeta_pp_1,CMS_dNdeta_pp_2}, indicating an 
efficiency of $(99.5\pm0.5)\%$. Despite larger particle multiplicities 
in PbPb collisions, the occupancy of the pixel detector is still smaller 
than 1\% owing to its high granularity (whereas in the strip detector it 
reaches 20\%), and the pixel detector exhibits the same excellent 
perfomance in PbPb as in pp collisions.

\begin{figure}
\begin{center}
\includegraphics[width=.45\textwidth]{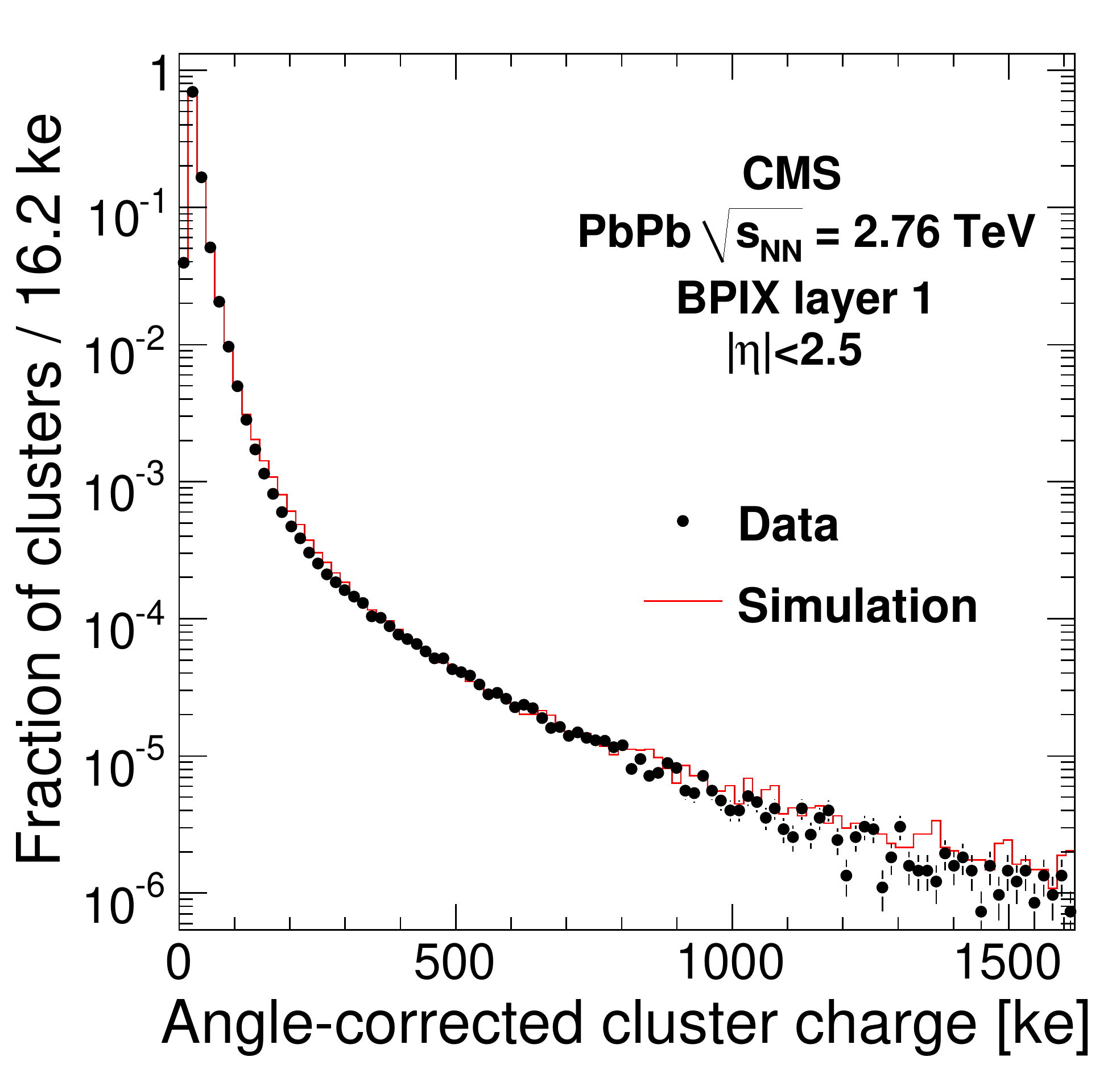}
\includegraphics[width=.45\textwidth]{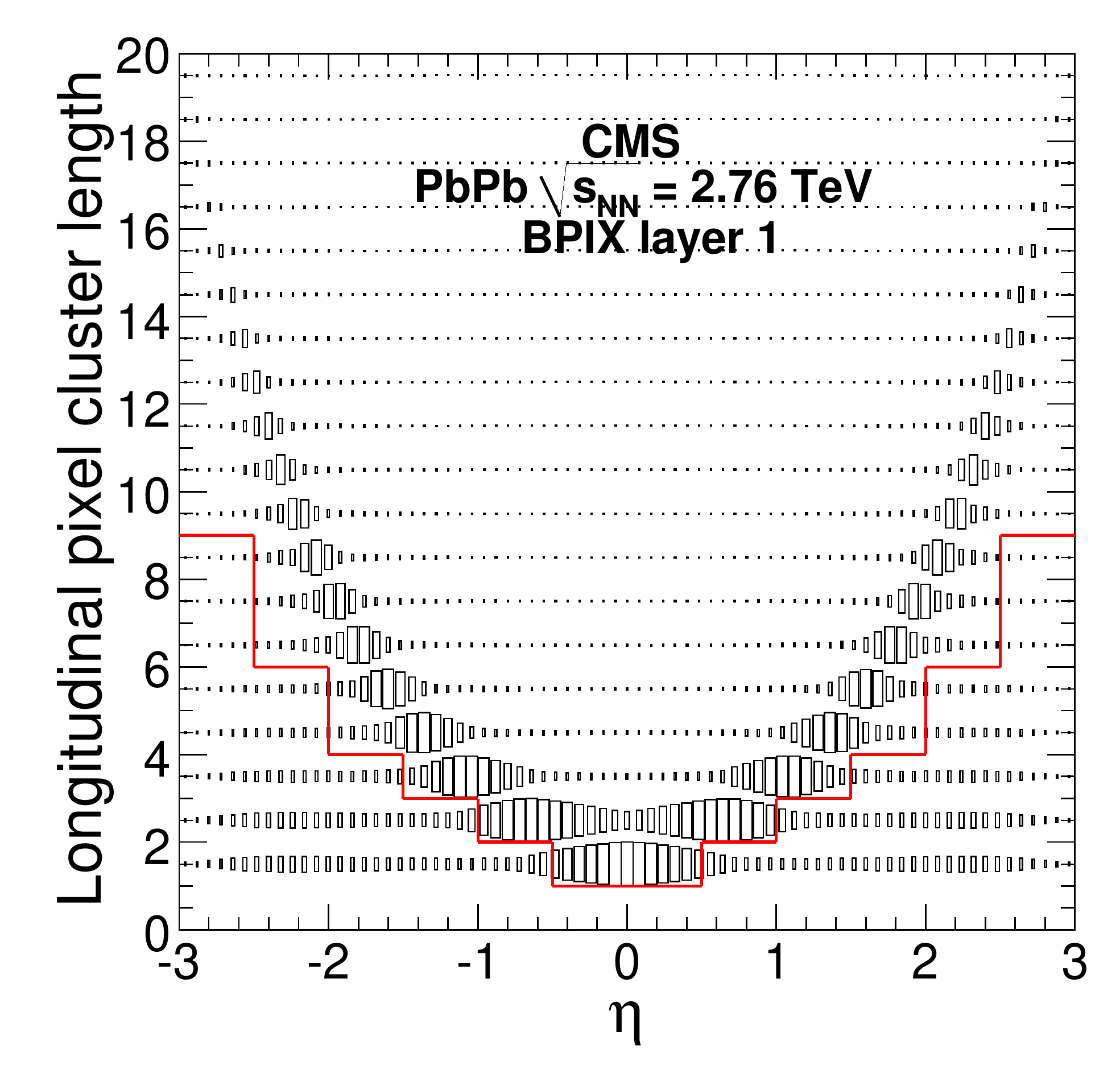}
\caption{{\it Left}: Distribution of the angle-corrected pixel-cluster 
charge in units of equivalent kilo-electrons from 2.76 TeV PbPb data 
and simulation. {\it Right}: Pixel-cluster length along the beam 
direction in units of pixel cells for hits from the first layer of the 
BPIX, as a function of $\eta$ after the event selection. The solid red 
line shows the selection on the minimum cluster length used in the 
analysis.}
\label{smiley}
\end{center}
\end{figure}

The hit-counting measurement method is based on the correlation between 
the cluster length in $z$ and the pseudorapidity of a particle 
originating from the interaction point. It measures the primary charged 
hadron multiplicity distributions using the occupancy of a certain layer 
of the pixel detector by counting the reconstructed hits. The 
hit-counting method gives three largely independent measurements for the 
three barrel layers. A similar method was used by the PHOBOS experiment 
at RHIC \cite{Back:2001bq} and also by CMS for earlier pp analyses 
\cite{CMS_dNdeta_pp_1,CMS_dNdeta_pp_2}. One of the disadvantages of the 
method is the strong reliance on detector simulation for correction 
factors. Therefore, the detector simulation was extensively studied and 
carefully compared to data. The simulation was found to give a very good 
description of the data in all observables related to detector 
performance. Such an example can be seen in the left panel of 
Fig.~\ref{smiley}, which shows the distribution of the pixel-cluster 
charge in the first BPIX layer from data and simulation. Clusters were 
selected according to the cluster-size selection described in 
Section~\ref{sec:clusterCounting_method}, and the cluster charge was 
normalised by the impact angle estimated from the cluster location and 
vertex position. The simulation describes the data well over six orders 
of magnitude.

\subsection{Primary vertex reconstruction using clusters}
\label{sec:cluster_vertexing}

There is a linear relationship between the length of a pixel-cluster 
along the beam direction and the $z$ position of the cluster. 
Thus, hits from primary tracks leave a characteristic V-shaped pattern 
in the plane of cluster size versus $z$ position. Nonprimary hits 
(e.g.\ due to secondary particles or nuclear interactions) fall mostly 
outside this V-shaped region. Thus, a V-shaped band is used to scan the 
$z$ axis; the $z$ position with the largest number of associated 
clusters is used as the vertex $z$ position. The vertex $z$ position is 
thus obtained by maximizing the consistency of the pixel-cluster 
lengths and global $z$ positions with a primary vertex hypothesis.

\subsection{Cluster selection}
\label{sec:clusterCounting_method}

Particles travelling from the primary vertex at a small angle with 
respect to the beam axis produce larger clusters in the BPIX layers than 
those at large angles. The cluster size is proportional to 
$|\sinh\eta|$, where the pseudorapidity $\eta$ of the cluster is 
computed with respect to the reconstructed vertex (right panel of 
Fig.~\ref{smiley}). Particles from background processes (decays in 
flight, nuclear interactions, etc.) often have smaller clusters than 
those produced in the primary interaction, since their crossing angle is 
not correlated to the $\eta$ of the hit. Thus, a large fraction of 
clusters from background processes (refered to as background clusters) 
can be rejected by a selection based on the cluster size variable. The 
selection is defined in $\eta$ bins (shown as a red line in the right 
panel of Fig.~\ref{smiley}).

\subsection{Corrections}

Not all of the background clusters are removed by the cluster selection 
described above since they can occasionally mimic the length of clusters 
generated by primary particles. The correction factor $\chi(\eta,M)$ is 
defined as the ratio of the number of selected clusters in the data 
to the number of primary charged hadrons at a pseudorapidity $\eta$ 
and a given cluster multiplicity $M$. It is calculated from simulation 
using:

\begin{equation}
\chi(\eta,M) = 
\frac{N^{\rm MC}_{\rm hit}(\eta,M)}{N^{\rm MC}_{\rm hadron}(\eta,M)},
\end{equation}

where $M$ denotes the total number of clusters passing the cluster 
selection, $N^{\rm MC}_{\rm hit}(\eta,M)$ is the number of selected 
clusters, and $N^{\rm MC}_{\rm hadron}(\eta,M)$ the number of primary 
charged hadrons in the simulation. This correction factor is used to 
convert the measured $N_{\rm hit}(\eta,M)$ pixel-cluster distributions 
from data into the corresponding primary charged hadron distributions, 
$N_{\rm hadron}(\eta,M)$.

The $\chi(\eta,M)$ correction factor is only weakly dependent on the 
physical process producing the hadrons, since it mainly contains 
information on the detector geometry. In a perfectly hermetic and 
$100\%$ efficient detector, $\chi(\eta,M)$ would be slightly above 
unity, as not only the primary, but also the secondary particles can 
generate hits. For detectors covering a limited solid angle, its values 
will be between $0$ and $1$. For very large multiplicities, 
$\chi(\eta,M)$ may decrease with increasing $M$ because of the more 
significant occupancy (provided the primary/secondary ratio stays 
roughly constant). However, the occupancy of the silicon pixel layers is 
observed to be small and no apparent decrease of $\chi(\eta,M)$ is 
visible with increasing centrality. The $\chi$ correction increases with 
increasing distance from the interaction point, because layers further 
from the primary vertex are hit by more decay products, as well as by 
more secondaries from nuclear interactions.

The correction factor in the first layer is in the range 1.0--1.2, 
while its average value in the 0--10\% centrality bin is $1.1$, $1.2$, 
and $1.3$ for the first, second, and third layers, respectively.

The pseudorapidity distribution of charged particles, for a fixed $M$, 
is calculated from the measured $N_{\rm hit}(\eta,M)$ distribution, 
correcting for the hit/primary charged hadron ratio and normalising it 
to the number of events with multiplicity $M$ passing the event 
selection, $N_{\rm selected}(M)$:

\begin{equation}
\label{eq:mult_dep_dNdeta}
\frac{dN_{\rm ch}}{{d\eta}} (\eta,M) = 
\frac{1}{{\Delta\eta\ \chi(\eta,M)}} 
\frac{{N_{\rm hit}(\eta,M)}}{{N_{\rm selected}(M)}},
\end{equation} 

where $\Delta\eta$ is the width of the $\eta$ bin. 

The event selection efficiency for a given multiplicity is determined as 
the ratio of the number of MC events with multiplicity $M$ which pass 
the event selection criteria $N_{\rm selected}^{\rm MC}(M)$ to the total 
number $N_{\rm tot}^{\rm MC}(M)$ generated with multiplicity $M$: 
$\epsilon(M) = N_{\rm selected}^{\rm MC}(M)/N_{\rm tot}^{\rm MC}(M)$.

To measure the final, multiplicity-independent pseudorapidity 
distribution, the multiplicity-dependent distributions derived from 
Eq.~\ref{eq:mult_dep_dNdeta} are weighted by the event selection 
efficiency $\epsilon(M)$ and then summed over $M$:

\begin{equation}
\label{eq:summation}
\frac{dN_{\rm ch}}{d\eta}(\eta) = \frac{{\sum_M N_{\rm selected}(M) 
{\frac{1}{\epsilon(M)} {\frac{dN_{\rm ch}}{d\eta}}(\eta,M)}}}{{\sum_M 
{N_{\rm selected}(M)\frac{1}{\epsilon(M)}}}}.
\end{equation}

Because a reconstructed event vertex is required as part of the event 
selection, the sum is over $M>0$.

\section{Tracklet method and corrections}
\label{chap:trackletMethod}

Tracklets are two-hit combinations in different layers of the BPIX that 
are consistent with a particle originating from the primary vertex. The 
tracklet analysis makes use of the correlation between hit positions: 
pairs of hits produced by the same charged particle have only small 
differences in the pseudorapidity ($\Delta\eta$) and the azimuthal angle 
($\Delta\phi$) with respect to the primary vertex.

\subsection{Primary vertex reconstruction using tracklets}
\label{sec:tracklet_vertexing}

In this method a tracklet-based vertex finder is used. In the first 
step, a hit from the first BPIX layer is selected and a matching hit is 
sought. If the magnitude of the difference in azimuthal angle 
($\Delta\phi$) between the two hits is smaller than 0.08, the pair is 
saved as a proto-tracklet. This procedure is repeated for each 
first-layer hit to get a collection of proto-tracklets. For each 
proto-tracklet, the expected longitudinal vertex position is found 
using:

\begin{equation}
z = z_1 - r_1 (z_2 - z_1) / (r_2 - r_1),
\end{equation}

where $z_{1(2)}$ is the $z$ position of the first (second) layer hit, 
and $r_{1(2)}$ is its radius. The calculated $z$ positions are saved as 
vertex candidates. The second step is to determine the primary vertex 
from the vertex candidates. If the magnitude of the difference between 
the $z$ positions of any two vertex candidates is less than 0.14~cm, 
they are combined as a vertex candidate cluster. Finally, the vertex 
candidate cluster with the highest number of vertex candidates is 
selected as the primary vertex. The final vertex $z$ position is 
determined by the average $z$ position of the vertex candidates in the 
cluster.

\subsection{Tracklet reconstruction}
\label{sect:trackletReconstruction}

All three barrel layers of the pixel detector are used in pairs: 
1st+2nd, 1st+3rd, and 2nd+3rd. The differences in 
pseudorapidity and azimuthal angle, as well as the
two-dimensional separation $\Delta R=\sqrt { (\Delta\eta)^2 + 
(\Delta\phi)^2 }$ between the two hits 
of a tracklet, are important for characterising the tracklet.

\label{sect:trackletReconstruction}

Tracklets are reconstructed in three steps:

\begin{enumerate}

\item For each reconstructed hit, the pseudorapidity is calculated using 
the primary vertex location. Hits that pass the cluster size selection 
(as described in Section~\ref{sec:clusterCounting_method}) are kept for 
further analysis.

\item Starting with a reconstructed hit in the $a^{\rm th}$ layer and 
looping over the reconstructed hits in the $b^{\rm th}$ layer (with 
$b>a$), all possible combinations with $|\Delta R|<0.5$ are saved as 
proto-tracklets.

\item Proto-tracklets are sorted in $\Delta R$. If a $b^{\rm th}$-layer 
hit is matched more than once, the proto-tracklet with the smallest 
$\Delta R$ is kept. The selected proto-tracklets are the final 
reconstructed tracklets.

\end{enumerate}

In addition to primary charged particles, the set of tracklets also 
include contributions from secondary interactions in the beampipe, 
particles from weak decays, and combinatorial background.

The combinatorial background tracklets are defined as combinations 
from secondary hits and hits from different primary tracks. The 
background fraction is largely suppressed by the $\Delta R$ ordering 
and the selection of tracklets (described in the next section). The 
tracklets from secondary particles are suppressed, and the correction 
for the remaining contribution relies on simulation.

\subsection{Combinatorial and secondary particle background}
\label{sect:combinatorialBackground} 

Background tracklets can be created from incorrectly associated hits. 
The $\Delta\eta$ and $\Delta\phi$ of a tracklet are very useful 
quantities for the separation of signal and combinatorial background 
tracklets. Because of the absence of magnetic field, selecting the best 
proto-tracklet with the smallest $\Delta R$ provides a powerful way to 
reject combinatorial background: signal proto-tracklets exhibit a 
correlation peak around $\Delta R = 0$, while the background component 
extends to large $\Delta R$.

\begin{figure}[htb]
\begin{center}
\resizebox{0.45\textwidth}{!}{\includegraphics{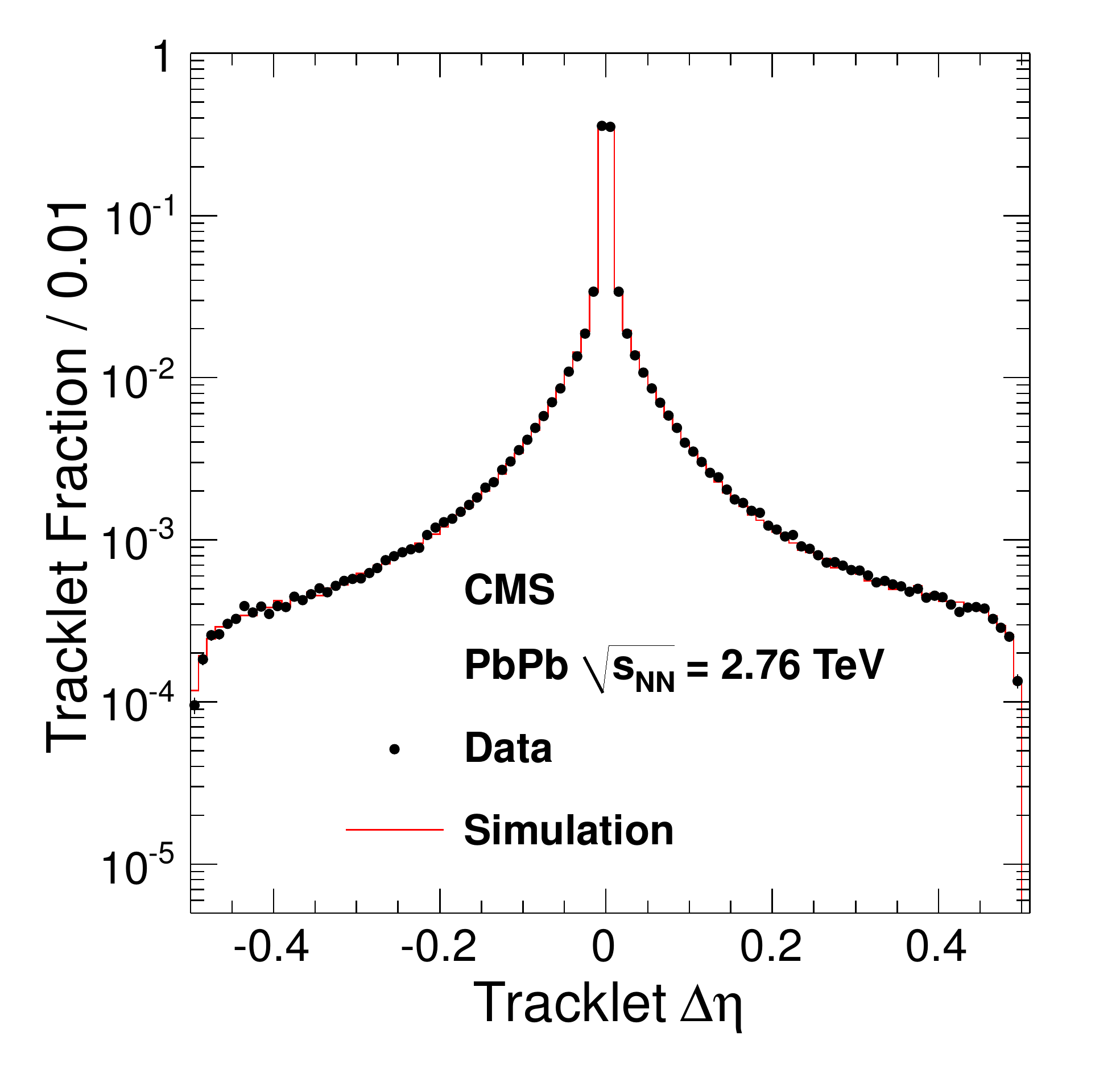}}
\resizebox{0.45\textwidth}{!}{\includegraphics{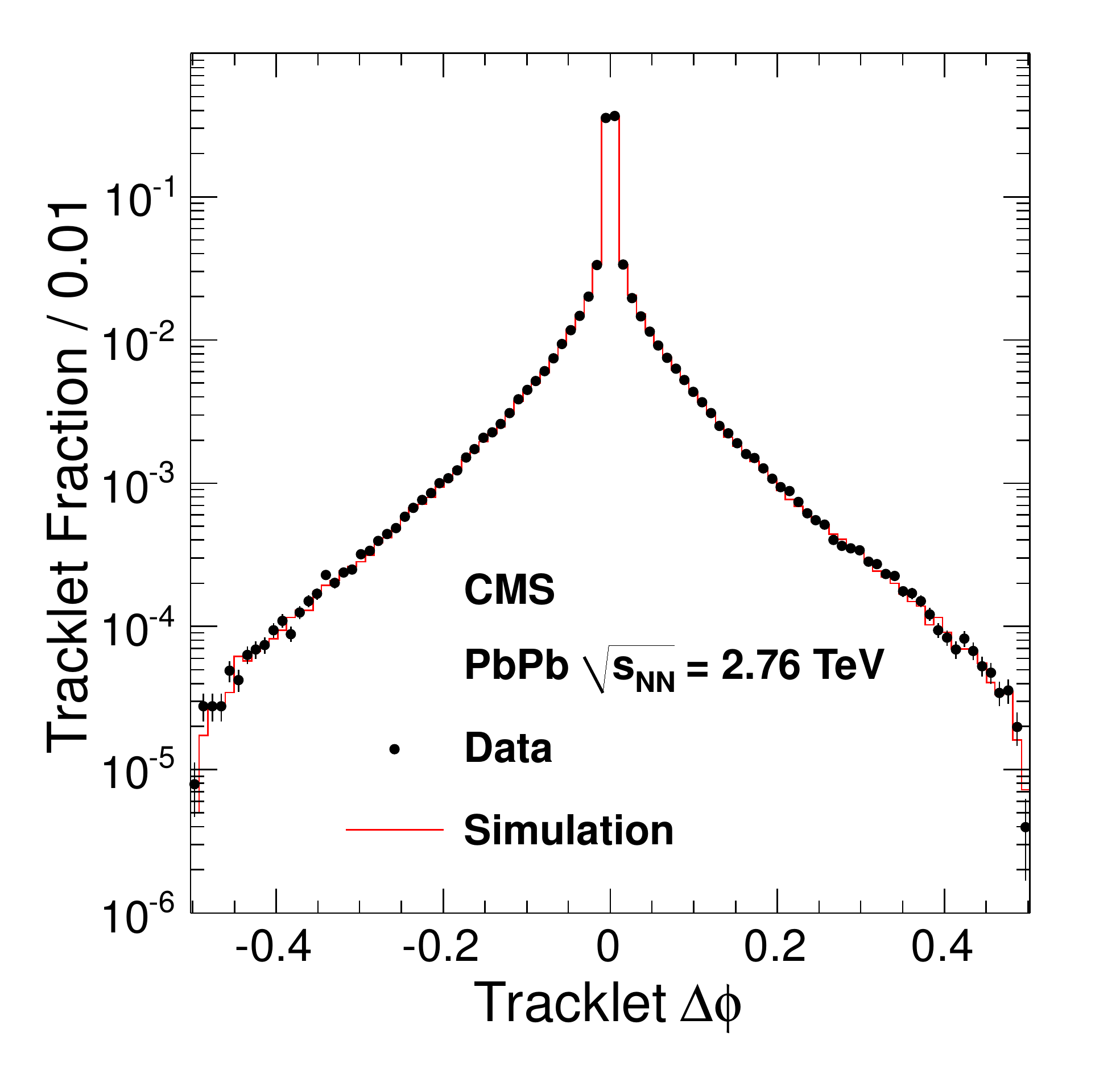}}
\caption{\label{fig:CleanTracklet12} The ({\it left}) 
$\Delta\eta$ and ({\it right}) $\Delta\phi$ distributions for 
reconstructed tracklets in minimum-bias collisions from the first and 
the second pixel layers in data and simulation.}
\end{center}
\end{figure}

The $\Delta\eta$ and $\Delta\phi$ distributions of selected tracklets 
from minimum-bias collisions in data and simulation are shown in 
Fig.~\ref{fig:CleanTracklet12} for combinations in the first and second 
pixel layers. The signal peaks at $\Delta\eta$ and $\Delta\phi=0$ are 
clearly visible. Data and simulation show agreement over several orders 
of magnitude.

The effect of secondary hits on the agreement between data and 
simulation seen in Fig.~\ref{fig:CleanTracklet12} was tested by adding 
random hits to simulated events. The simulated tracklet spectra were 
found to be distorted even by a few percent of random hits, spoiling the 
agreement between data and simulation. Given the very good agreement in 
the tracklet spectra, the fraction $\beta$ of combinatorial background 
tracklets can therefore be reliably obtained from simulation. The value 
of $\beta$ is in the range 0--15\%, depending on the multiplicity of the 
event, the pseudorapidity of the tracklets, and the $z$ position of the 
event vertex. The number of background-subtracted tracklets $N_{\rm 
tracklet}$ is determined from the raw number of tracklets in data 
$N_{\rm tracklet}^{\rm raw}$ using: $N_{\rm tracklet} = (1-\beta)\times 
N_{\rm tracklet}^{\rm raw}$.

\subsection{Efficiency and acceptance correction}

To calculate the number of hadrons from the number of tracklets, an 
efficiency correction must be applied. The correction factor 
$\alpha(M,\eta,z_{\rm v})$ for the tracklet reconstruction efficiency is 
defined as

\begin{equation}
 \alpha(M,\eta,z_{\rm v})=\frac{N_{\rm hadron}^{\rm truth}(M,\eta,z_{\rm 
v})}{[1-\beta(M,\eta,z_{\rm v})] N_{\rm tracklet}^{\rm raw, 
MC}(M,\eta,z_{\rm v})},
\end{equation}

where $z_{\rm v}$ is the $z$ position of the vertex, $N_{\rm 
hadron}^{\rm truth}(M,\eta,z_{\rm v})$ is the true number of charged 
hadrons in the simulated sample, and $N_{\rm tracklet}^{\rm raw, MC}$ is 
the raw number of selected tracklets in the MC sample. The factor 
$\alpha(M,\eta,z_{\rm v})$ is used to calculate the charged hadron 
spectra from the measured background-subtracted tracklets. Typical 
values of $\alpha$ are less than 1.15 because of the high 
hit-reconstruction efficiency in the BPIX. At larger pseudorapidity, the 
correction factor increases, owning to the reduced acceptance. The size 
of the acceptance correction also depends on the position of the primary 
vertex.

The pseudorapidity distribution of charged hadrons for a given 
multiplicity $M$ is obtained from the measured number of tracklets 
($N_{\rm tracklet}^{\rm raw})$, the background fraction ($\beta$), the 
efficiency and acceptance correction ($\alpha$), and the normalisation 
to the number of selected events:

\begin{eqnarray}
\frac{dN_{\rm ch}}{d\eta}(\eta,M) =
\frac{\sum_{z_{\rm v}}{\alpha(M,\eta,z_{\rm v})[1-\beta(M,\eta,z_{\rm 
v})]N_{\rm tracklet}^{\rm raw}(M,\eta,z_{\rm v})}}{\Delta\eta 
\ N_{\rm selected}(M)},
\end{eqnarray}

where $\Delta\eta$ is the width of the $\eta$ bin and $N_{\rm 
selected}(M)$ is the number of selected events used in each multiplicity 
bin. The $\alpha(1-\beta)$ correction has a typical value larger than 
$0.85$. For the final $dN_{\rm ch}/d\eta$ distribution the 
multiplicity-dependent results are weighted by the event selection 
efficiency $\epsilon(M)$ and then summed as in Eq.~\ref{eq:summation}.

\section{Systematic uncertainties}
\label{systerr_summary}

A summary of systematic uncertainties affected the measurement of 
$dN_{\rm ch}/d\eta$ for the two analysis methods is given in 
Table~\ref{tab:syst-table}.

The results from different BPIX layers (or layer combinations) from 
either of the measurement methods differ by less than $2\%$, and thus 
they are averaged using the arithmetic mean. The uncertainties of the 
averaged results from the hit-counting and tracklet methods are 
dominately systematic and largely correlated. Therefore, the two results 
are averaged using equal weights. Since the difference between the 
measurements from the two methods is smaller than $1\%$, the weighting 
procedure has very little effect on the numerical value of the final 
result.

The uncertainties on the final average result are computed as follows. 
All the systematic uncertainties listed in Table~\ref{tab:syst-table} 
are correlated between the two methods, except those associated with the 
efficiency of reconstruction and misalignment. 
The correlated, $(s_{\rm h.c.})_j$ and $(s_{\rm tracklet})_j$, and the 
uncorrelated, $(\sigma_{\rm h.c.})_j$ and $(\sigma_{\rm tracklet})_j$ 
uncertainties of the hit-counting and tracklet methods (indexed by $j$) 
are summed in quadrature: 
\begin{equation}
\bar{s}=\sqrt{\Sigma_j[(s_{\rm h.c.})_j+(s_{\rm tracklet})_j]^2}\big/2 
\ \ {\rm and }\ \ \bar{\sigma}=\sqrt{(\Sigma_j[(\sigma_{\rm 
h.c.})_j^2+(\sigma_{\rm tracklet})_j^2]}\big/2,
\end{equation}

resulting in $\bar{s}$ and $\bar{\sigma}$ correlated and uncorrelated 
uncertainties of the average, respectively. The total systematic 
uncertainty is then $\bar{\sigma}_{\rm tot}=\sqrt{\bar{\sigma}^2 + 
\bar{s}^2}$.

In this paper the distributions of three observables are reported: 
$dN_{\rm ch}/d\eta|_{\eta=0}$ as a function of centrality class, 
$(dN_{\rm ch}/d\eta)/(N_{\rm part}/2)$ as a function of $\eta$, and 
$(dN_{\rm ch}/d\eta|_{\eta=0})/(N_{\rm part}/2)$ as a function of 
$N_{\rm part}$.

The systematic uncertainties affecting the slope and those affecting the 
absolute scale of \\$dN_{\rm ch}/d\eta|_{\eta=0}$ and $(dN_{\rm 
ch}/d\eta|_{\eta=0})/(N_{\rm part}/2)$ measurements are determined 
separately. Systematic uncertainty sources affecting the slope are those 
on the centrality and the Glauber calculation of $N_{\rm part}$; all 
other sources affect the scale. The results are presented in 
Section~\ref{sec:Results} with these two uncertainties shown separately.

The slope of $dN_{\rm ch}/d\eta|_{\eta=0}$ as a function of centrality 
is only affected by the uncertainty on the determination of the 
centrality bins. Both the slope and the absolute scale of the $N_{\rm 
part}$-normalised distributions are affected by the uncertainty on 
$N_{\rm part}$ from the Glauber calculation, given in 
Table~\ref{table:npart}. These contributions are computed by 
transforming the uncertainty in $N_{\rm part}$ into an uncertainty on 
the $N_{\rm part}$-normalised hadron density distribution using the 
derivative of the measured $(dN_{\rm ch}/d\eta)/(N_{\rm part}/2)$ 
distributions as a function of $N_{\rm part}$.

\begin{table}[htb]
\centering
\caption{\label{table:systematics}Summary of systematic uncertainties on 
the $dN_{\rm ch}/d\eta$ measurements and their sum for the two analysis 
methods.}
 \label{tab:syst-table}
\begin{tabular}{lcccc}
\hline
\hline
 Source                                          & Hit-counting [\%] & Tracklet [\%]\\
\hline
 Centrality (0--5\% to 85--90\%)                 & 0.5--15.6     & 0.5--15.6               \\
 Pixel hit efficiency                            & 0.5           & 1.0                \\
 Tracklet and cluster selection                  & 3.0           & 0.5                \\
 Acceptance uncertainty                          & 1.5           & 1.5                \\
 Correction for secondary particles              & 2.0           & 1.0                \\
 Pixel-cluster splitting                         & 1.0           & 0.4                \\
 Reconstruction efficiency                       & -             & 1.9                \\
 Misalignment                                    & -             & 1.0                \\
 Random hits                                     & 1.0           & 0.2                \\
\hline
 Total uncorrelated uncertainties               & -              & 2.1                \\
 Total uncertainties                            & 4.2--16.2      & 3.1--15.9                \\
\hline
\hline
\end{tabular} 
\end{table}

The systematic uncertainties affecting the measurements are as follows.

$\bullet$ {\bf Centrality}: The determination of the centrality bins as 
a percentage of the total hadronic cross section relies on the hadronic 
event selection efficiency ($99\pm1$)\%, UPC contamination 
($1\pm0.5$)\%, and the percentile binning of the centrality variable. 
Thus, the uncertainty in the event selection cross section causes 
uncertainty in the centrality binning by moving the bin boundaries, 
shifting the event population in each centrality bin. The effect of this 
centrality uncertainty on the final results was studied by repeating 
the analysis using various centrality tables (derived from the various 
trigger efficiencies allowed by the uncertainty on the efficiency). 
These studies indicate that the uncertainty of the $dN_{\rm ch}/d\eta$ 
values ranges from 0.5\% for the 0--5\% centrality bin to 15.6\% for 
the 85--90\% centrality bin.

$\bullet$ {\bf Pixel hit efficiency}: The efficiency of the pixel layers 
is larger than $99\%$, measured from pp data \cite{CMS_dNdeta_pp_1}. The 
pixel detector has low occupancy even in central heavy-ion collisions 
because of its fine segmentation. Therefore, the same pixel hit 
efficiency and uncertainty measured in pp collisions are used here. The 
pixel hit efficiency affects tracklets more, since two layers are 
required; a 0.5\% inefficiency per pixel layer leads to a 1\% 
inefficiency for tracklet finding.

$\bullet$ {\bf Tracklet and cluster selection}: Varying the cluster 
selection requirements (pixel-cluster length selection) and tracklets 
selection (requirement on $\Delta R$) is used to estimate the 
uncertainty due to cluster and tracklet selection. The cluster 
selections were changed by one pixel unit in all $\eta$ bins and the 
$\Delta R$ selection by a factor of three. The observed differences in 
the final results (3\% and 0.5\%, respectively) are quoted as 
conservative systematic uncertainties.

$\bullet$ {\bf Acceptance uncertainty}: The positions of the BPIX 
modules are only slightly different in data and in simulation, but hits 
at the extreme edges of the BPIX are not used in the analysis, limiting 
the systematic uncertainty from this effect. The $\eta,\phi$ acceptance 
was estimated from data by using the endpoints of tracklets to map the 
active surface of the BPIX layers. From this study, the acceptance 
correction is estimated to be 1\% in pp collisions. In PbPb collisions 
(due to the longer luminous region in the beam direction) this 
uncertainty was increased to 1.5\%. No correction is applied, but the 
effect is included in the systematic uncertainty.

$\bullet$ {\bf Corrections due to hits from secondary particles}: The 
sensitivity of the correction factors applied to remove hits caused by 
secondary particles was tested using two largely different event 
generators: {\sc ampt} and {\sc hydjet} \cite{Hydjet}. The relative 
fraction of strange particle production differs by 60\% in the two 
generators, but the effect on the correction factor was found to be only 
$2\%$ for the case of the hit-counting analysis. The tracklet analysis 
is less sensitive to secondaries.

$\bullet$ {\bf Pixel cluster splitting}: The relative fraction of split 
clusters was estimated from the cluster-cluster distance distribution. 
This study shows that the number of split clusters in data does not 
exceed that in simulation by more than 0.5--0.7\%. No correction is 
applied for this effect in the analyses, but a conservative systematic 
uncertainty (1\% and 0.4\%, respectively, for the hit-counting and 
tracklet analyses) is assigned.

$\bullet$ {\bf Efficiency of tracklet reconstruction}: The uncertainties 
in the simulation of the $p_{\rm T}$ and multiplicity ($M$) 
distributions influence the determination of the tracklet reconstruction 
efficiency. The uncertainty (1.9\%) is estimated based on variations of 
these quantities within reasonable limits: $\langle p_{\rm T}\rangle$ 
was modified by 10\%, and the multiplicity variable was changed from 
using clusters to HF towers.

$\bullet$ {\bf Misalignment}: The hit-counting method is not sensitive 
to detector misalignments. The tracklet method has a sensitivity through 
the $\Delta R$ selection, which was studied by moving the reconstructed 
hit positions (the entire detector) by 0.3 mm in the simulation, while 
keeping the vertex position at the same place, giving a 1\% change in 
the final result. Since the $\Delta\phi$ and $\Delta\eta$ distributions 
and the correlation widths agree well, no significant misalignment is 
seen in the data.

$\bullet$ {\bf Random hits}: With the restrictive event selection 
criteria, the contamination from beam-gas (high-occupancy) events in the 
final data sample is negligible. The other potential source of 
background is the accidental overlap between beam-gas and PbPb 
collisions. A conservative systematic uncertainty of 1\% is assigned for 
the hit-counting analysis, which is more sensitive to this overlap 
than the tracklet analysis, for which 0.2\% is assigned.

\section{Results}
\label{sec:Results}

The hit-counting and tracklet $dN_{\rm ch}/d\eta$ results are in 
good agreement; their average difference is smaller than 1\%. Their 
individual results are averaged as described in 
Section~\ref{systerr_summary}, and these averages are presented as the 
final results.

\begin{figure}
\begin{center}
\includegraphics[width=.45\textwidth]{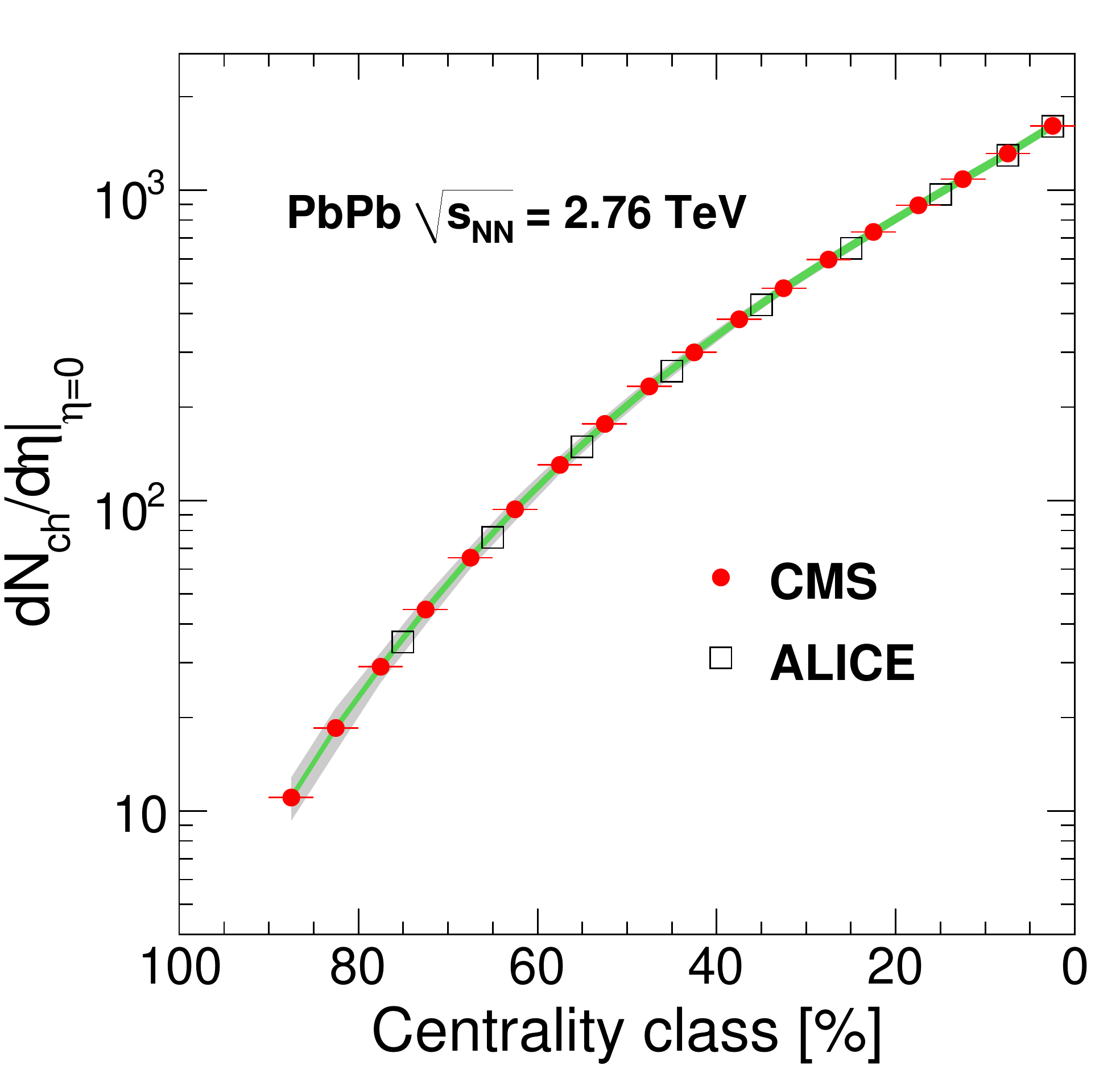}
\includegraphics[width=.45\textwidth]{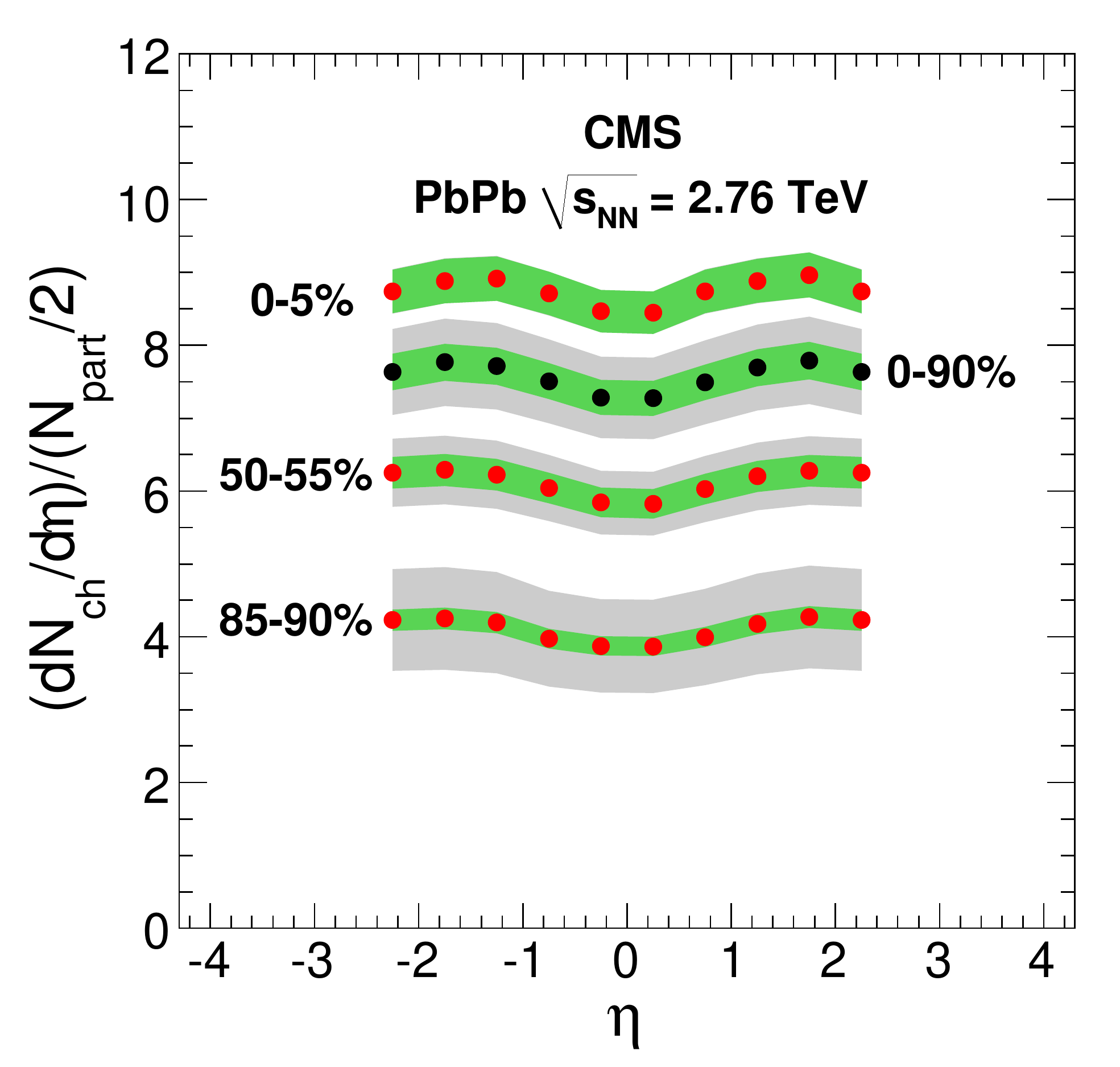}
\caption{
{\it Left}: $dN_{\rm ch}/d\eta|_{\eta=0}$ as a function of 
centrality class in 2.76~TeV PbPb collisions from this 
experiment (solid circles) and from ALICE (open squares) 
\cite{ALICE_dNdeta}. The inner green band shows the measurement
uncertainties affecting the scale of the measured distribution from 
this analysis, while
the outer grey band shows the full systematic uncertainty, i.e.\ 
affecting both the scale and the slope.
{\it Right}: Measured $dN_{\rm ch}/d\eta/(N_{\rm 
part}/2)$ distributions from this analysis as a function of $\eta$ in 
various centrality bins.}
\label{fig:dNdeta}
\end{center}
\end{figure}

The left panel of Fig.~\ref{fig:dNdeta} presents the measured $dN_{\rm 
ch}/d\eta|_{\eta=0}$ values as a function of centrality. The statistical 
uncertainties are negligible, while the systematic uncertainties are 
shown as two bands. The inner green band shows the measurement 
uncertainties affecting the scale of the measured distribution, while 
the outer grey band shows the full systematic uncertainty, 
i.e.\ affecting both the scale and the slope. Details on the calculation 
of the uncertainty bands are given in Section~\ref{systerr_summary}. The 
charged hadron density for the 5\% most-central events (0--5\% 
centrality bin) is measured to be $dN_{\rm ch}/d\eta|_{\eta=0} = 1612\pm 
55\text{ }(syst.)$. These results are consistent with those of ALICE 
\cite{ALICE_dNdeta} within the uncertainties, as shown in 
Fig.~\ref{fig:dNdeta} (left). The error bars of the ALICE points in the 
figure show the total statistical and systematic uncertainties. The CMS 
measurements cover the centrality range of 0--90\%, extending the ALICE 
results (0--80\%) to more-peripheral collisions.

In order to compare bulk particle production for different colliding 
nuclei and at different energies, the charged-hadron density is divided 
by the average number of participating nucleon pairs, $N_{\rm part}/2$, 
determined for each centrality bin. The $N_{\rm part}$ values are 
obtained using the Glauber calculation, by classifying events according 
to their impact parameter, without reference to a specific particle 
production model (Table~\ref{table:npart}).

The measured $(dN_{\rm ch}/d\eta)/(N_{\rm part}/2)$ distributions as a 
function of $\eta$ in various centrality bins are shown in the right 
panel of Fig.~\ref{fig:dNdeta}. The uncertainty bands of these 
distributions also include the Glauber uncertainty on $N_{\rm part}$. 
The $\eta$ dependence of the results is weak, varying by less than 10\% 
over the $\eta$ range. The slight dip at $\eta = 0$ is a trivial 
kinematic effect (Jacobian) owing to the use of pseudorapidity ($\eta$) 
rather than rapidity ($y$).

\begin{figure}
\begin{center}
\includegraphics[width=.45\textwidth]{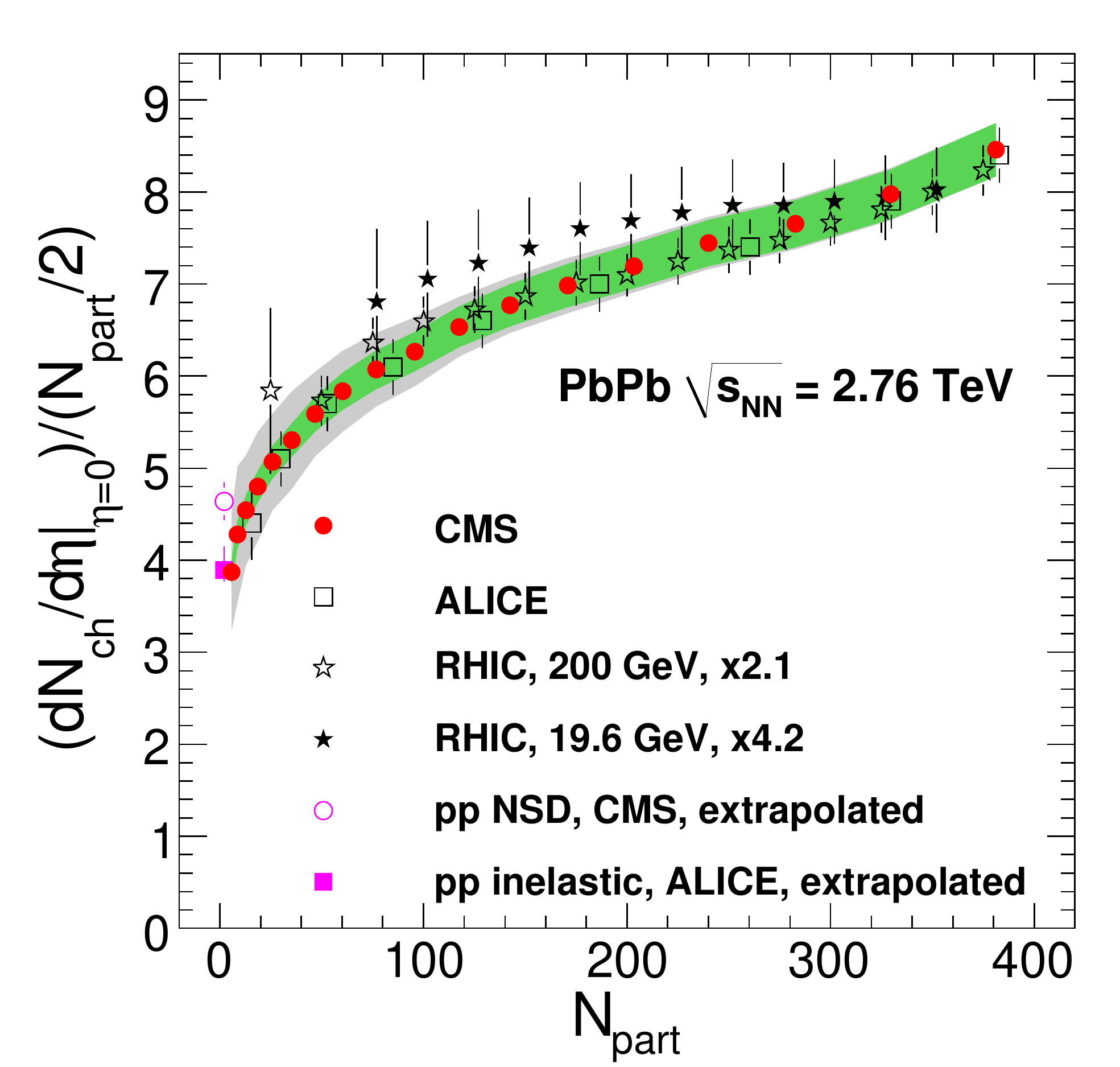}
\includegraphics[width=.45\textwidth]{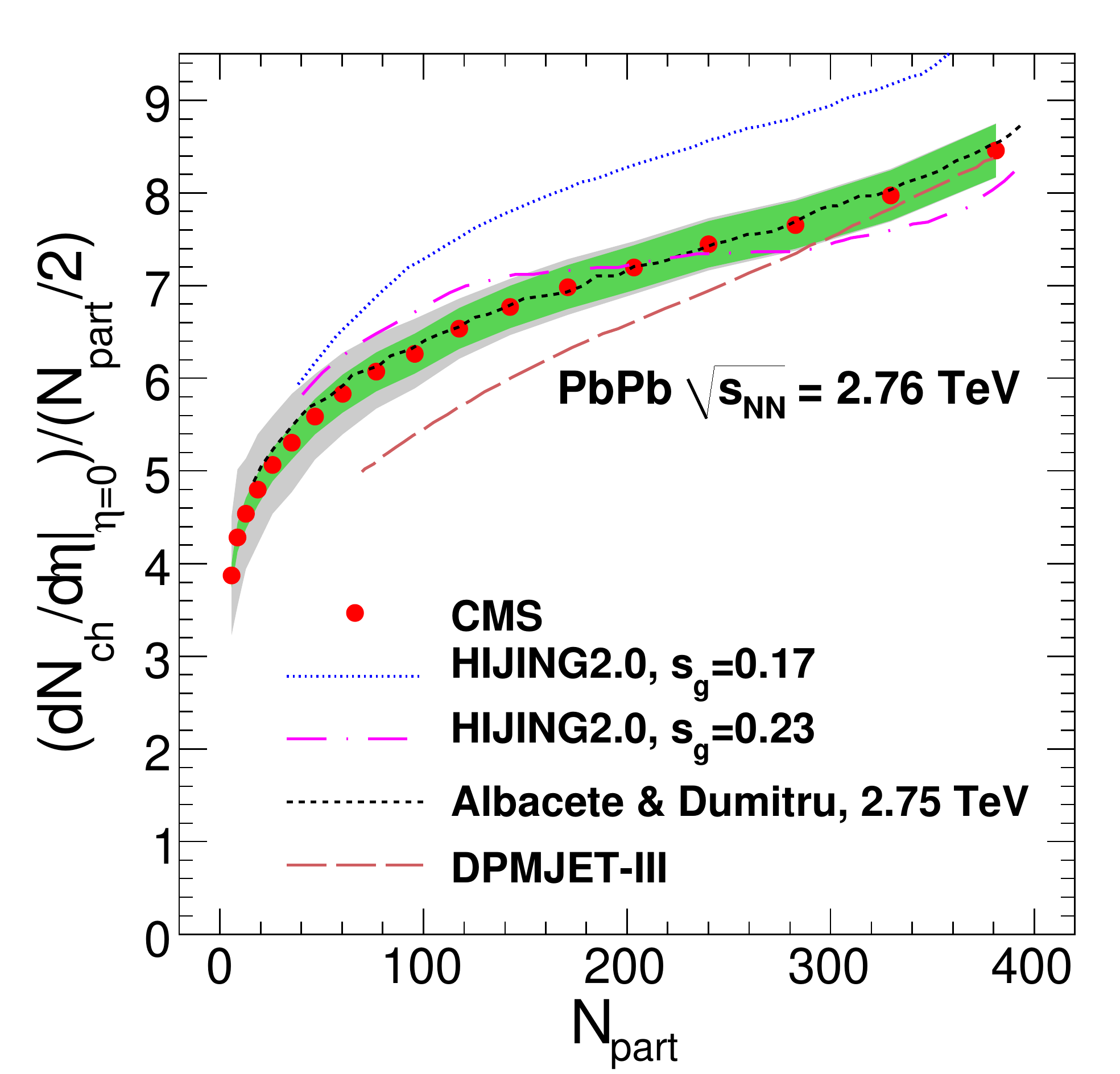}
\caption{ 
{\it Left}: Measured $(dN_{\rm ch}/d\eta|_{\eta=0})/(N_{\rm 
part}/2)$ as a function of the number of participants in 2.76~TeV 
PbPb collisions from this analysis and the ALICE experiment 
\cite{ALICE_dNdeta}, from RHIC \cite{RHIC_av} at 200 GeV and 19.6 GeV, 
and from extrapolated pp results from CMS \cite{CMS_dNdeta_pp_1} and 
ALICE \cite{ALICE_pp}. Systematic uncertainties affecting the scale of 
the measurements from this analysis are shown 
as inner green error bands and the total systematic uncertainties as an 
outer grey band, while the error bars indicate statistical uncertainties. 
The black stars are shifted slightly to the right for better 
visibility. The ALICE and the averaged RHIC results are 
from \cite{ALICE_dNdeta} and \cite{RHIC_av}, respectively.
{\it Right}: Results from this analysis are compared with model 
predictions of $(dN_{\rm ch}/d\eta|_{\eta=0})/(N_{\rm part}/2)$ as a 
function of the number of participants in 2.76 TeV PbPb collisions. The 
model predictions are taken from Refs.~\cite{HIJING2.0}, 
\cite{Albacete}, and \cite{DPMJET}.}
\label{fig:dNdeta_npart}
\end{center}
\end{figure}

The left panel of Fig.~\ref{fig:dNdeta_npart} presents the measured 
$(dN_{\rm ch}/d\eta|_{\eta=0})/(N_{\rm part}/2)$ as a function of 
$N_{\rm part}$. The statistical uncertainties on the CMS results are 
indicated by error bars (negligible), while the systematic uncertainties 
are shown as two bands. The inner green band shows the systematic 
uncertainty affecting the scale, while the outer grey band shows the 
full systematic uncertainty. The error bars on the ALICE 
\cite{ALICE_dNdeta} and the RHIC \cite{RHIC_av} points show the 
quadratic sum of the statistical and systematic uncertainties. The RHIC 
results are multiplied by numerical factors to match the 
$N_{\rm part}$-normalised multiplicity observed at the LHC for central 
collisions. The pp results shown in the figure are obtained from the 
measured non-single-diffractive (NSD) $dN_{\rm 
ch}/d\eta|_{\eta=0}=4.47\pm0.2$ (CMS) \cite{CMS_dNdeta_pp_1} and 
the inelastic $dN_{\rm ch}/d\eta|_{\eta=0}=3.77^{+0.26}_{-0.13}$ (ALICE) 
\cite{ALICE_pp} values at 2.36 TeV, using the $\sqrt{s}$ dependence of 
the charged hadron multiplicity density measured in NSD and inelastic 
collisions from Ref.~\cite{CMS_dNdeta_pp_2}. The error bars on the pp 
points show the total (statistical and systematic) uncertainties. The 
$N_{\rm part}$ values used for the normalisation by CMS and ALICE differ 
by less than $2\%$. Within the uncertainties, the $N_{\rm 
part}$-normalised hadron densities follow a similar dependence on 
centrality for all centre-of-mass energies, although the lower-energy 
collider data appear to have a flatter dependence on $N_{\rm part}$.

The phenomenological descriptions of particle production in nuclear 
collisions are often based on two-component models, combining 
contributions from perturbative QCD processes, i.e.\ (mini)jet 
fragmentation and soft interactions. The data are compared to three 
different approaches: (i) {\sc hijing} 2.0 \cite{HIJING2.0}, which 
basically scales (via the number of incoherent nucleon-nucleon 
collisions) the (semi)hard parton scatterings and fragmentation (Lund 
model \cite{lund}) implemented in {\sc pythia} after accounting for the 
``shadowing'' of the nuclear parton distribution functions; (ii) parton 
saturation approaches \cite{Albacete}, which model heavy-ion 
interactions as the collision of two dense multigluon wavefunctions with 
cross sections peaking at a semihard scale (saturation momentum of 
$\approx$2--3~\GeVc\ at the LHC) \cite{Gelis:2010nm,Kharzeev:2002pc}, 
followed by their fragmentation according to a simple parton-to-hadron 
local-duality prescription; and (iii) the {\sc dpmjet}-III MC program 
\cite{DPMJET}, based on the Regge-Gribov theory. This is an extension of 
the {\sc phojet} \cite{phojet} program in which interactions from soft 
degrees of freedom (Pomerons) can fuse in the dense initial state. They 
are extended consistently into the hard regime via ``hard'' or ``cut'' 
Pomerons, and then fragmented using the standard Lund model.

The measured $(dN_{\rm ch}/d\eta|_{\eta=0})/(N_{\rm part}/2)$ versus 
$N_{\rm part}$ distribution is compared to the various model predictions 
in the right panel of Fig.~\ref{fig:dNdeta_npart}. The two-component 
{\sc hijing} 2.0 model, which has been tuned to high-energy pp and 
central PbPb data, describes the general shape of the data. The {\sc 
hijing} model includes an impact-parameter-dependent gluon shadowing 
parameter $s_g$, which limits the rise of particle production with 
centrality. The magnitude of the particle production favours a 
relatively large $s_g=0.23$ value, although the shape of the 
centrality dependence prefers a smaller $s_g = 0.17$. The observed 
centrality dependence is well reproduced by the saturation model of 
Ref.~\cite{Albacete}. Both Refs.~\cite{HIJING2.0} and \cite{Albacete} 
were published knowing the result of ALICE \cite{ALICE_centraldNdeta} on 
the multiplicity of the 5\% most-central collisions, although previous 
saturation-based calculations (e.g.\ \cite{Armesto:2004ud}) predicted 
central charged hadron densities very similar to those finally measured. 
The {\sc dpmjet}-III model is capable of describing the charged hadron 
multiplicity in the most-central collisions, but shows a stronger rise 
with centrality than observed in the data. The measured particle 
densities provide basic constraints on the initial conditions of the 
quark-gluon plasma in any hydrodynamical approach employed to study PbPb 
interactions at the LHC~\cite{Hirano:2007gc}.

\begin{figure}[h]
\begin{center}
\includegraphics[width=.45\textwidth]{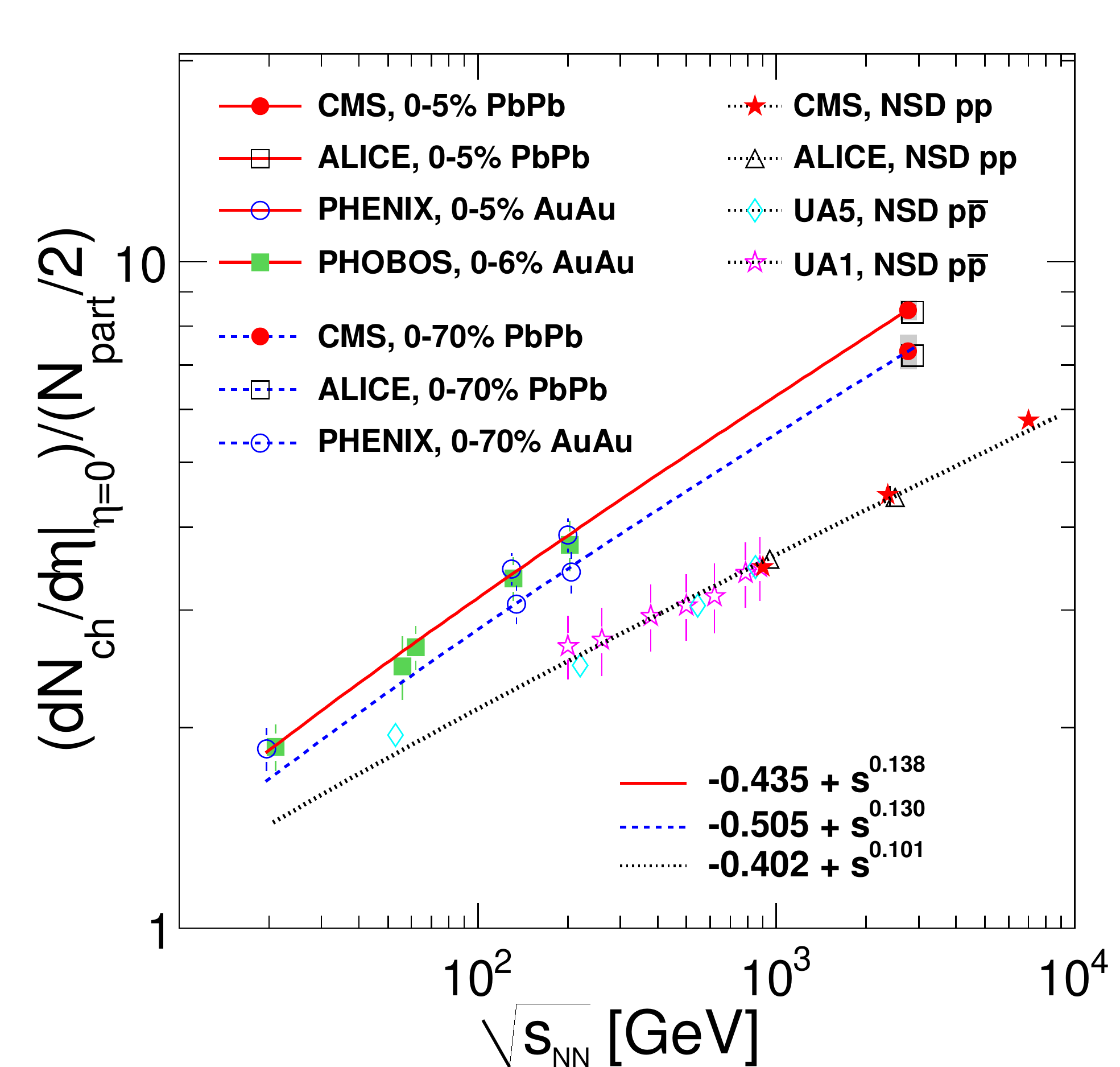}
\caption{Normalized charged hadron pseudorapidity density at $\eta=0$ as 
a function of centre-of-mass energy for the 0--5\% most-central 
nucleus-nucleus collisions (top set of points) and 0--70\% 
centrality (middle set), and for NSD pp collisions (bottom set). 
The fits to power-law functions are shown by lines. The grey band around 
the PbPb CMS points indicates the total systematic uncertainty. The 
statistical uncertainty is negligible. The 
error bars on the other points indicate statistical and 
systematic errors. The ALICE, PHENIX, and PHOBOS results (which are 
shifted slightly to the right for better visibility) are taken 
from Refs.~\cite{ALICE_dNdeta}, \cite{RHIC_av}, and \cite{dNdeta_PHOBOS}, 
respectively. The NSD pp results of CMS, ALICE, UA5, and UA1 are from 
Refs.~\cite{CMS_dNdeta_pp_1,CMS_dNdeta_pp_2}, \cite{ALICE_pp}, \cite{Alner:1986xu}, and \cite{Albajar:1989an}, respectively.}
\label{fig:sqrts}
\end{center}
\end{figure}

The collision-energy dependence of the measured $(dN_{\rm 
ch}/d\eta|_{\eta=0})/(N_{\rm part}/2)$ for 0--5\% and 0--70\% centrality 
from this analysis and from ALICE, PHENIX, and PHOBOS can be seen in 
Fig.~\ref{fig:sqrts}. The PHENIX and PHOBOS points are taken from 
Refs.~\cite{RHIC_av} and \cite{dNdeta_PHOBOS}, their error bars 
representing both the statistical and systematic uncertainties. 
Systematic uncertainties of the measurements from this analysis are 
shown as an error band, while the statistical uncertainties are 
negligible. The NSD pp results of CMS, ALICE, UA5, and UA1 are from 
Refs.~\cite{CMS_dNdeta_pp_1,CMS_dNdeta_pp_2}, \cite{ALICE_pp}, 
\cite{Alner:1986xu}, and \cite{Albajar:1989an}, respectively. The 
$N_{\rm part}$ values at different collision energies are different for 
a fixed centrality bin. When the $N_{\rm part}$ dependence of $(dN_{\rm 
ch}/d\eta)/(N_{\rm part}/2)$ from PHENIX and PHOBOS are used to 
extrapolate their $(dN_{\rm ch}/d\eta)/(N_{\rm part}/2)$ results shown 
in Fig.~\ref{fig:sqrts} to the $N_{\rm part}$ values appropriate for the 
LHC, they change by no more than 3\%. This correction is not applied 
in Fig.~\ref{fig:sqrts}. The normalised charged hadron densities shown 
in Fig.~\ref{fig:sqrts} are fit to a power-law function: $a+s_{_{\rm 
NN}}^n$. The fit returns the value $n=0.13$ for PbPb and $n=0.10$ for 
NSD pp collisions. The results of the fits are shown by the straight 
lines in Fig.~\ref{fig:sqrts}. These results provide additional 
constraints on the energy evolution of the saturation momentum in the 
proton and nuclei \cite{Gelis:2010nm,Kharzeev:2002pc}, as well as in 
general on the $p_{\rm T}$ cutoff between soft and hard dynamics used in 
the models of particle production in high-energy hadronic collisions.

\section{Summary}
\label{chap:conclusions}

A measurement of charged hadron multiplicity as a function of 
pseudorapidity and centrality in PbPb collisions at $\sqrt{s_{_{\rm 
NN}}}=2.76$ TeV has been reported. For the 5\% most-central collisions, 
a primary charged hadron density of $1612\pm55$ is measured, which 
represents an increase of a factor of 3 compared to similar measurements 
at RHIC energies. The $dN_{\rm ch}/d\eta$ distributions, measured over 
the range $|\eta|<2.5$, show weak $\eta$ dependence, the variation being 
less than 10\%. The $N_{\rm part}$-normalised multiplicity distributions 
from RHIC and the LHC have a similar dependence on centrality, although 
the lower-energy collider data has a somewhat flatter dependence. A 
parton saturation model describes well the observed centrality 
dependence. The collision-energy dependence of the measured hadron 
multiplicities at central rapidities is well modelled by a power-law 
function of the type $a+s_{_{\rm NN}}^{n}$. These results provide 
information on the parton structure of the nucleus and the proton and 
its evolution as a function of centre-of-mass energy. They also give 
additional constraints on the initial conditions in nucleus-nucleus 
collisions at LHC energies for hydrodynamical evolution studies of the 
strongly interacting produced system.

\section*{Acknowledgments}
\label{sect:acknowledge}

We wish to congratulate our colleagues in the CERN accelerator 
departments for the excellent performance of the LHC machine. We thank 
the technical and administrative staff at CERN and other CMS institutes, 
and acknowledge support from: FMSR (Austria); FNRS and FWO (Belgium); 
CNPq, CAPES, FAPERJ, and FAPESP (Brazil); MES (Bulgaria); CERN; CAS, 
MoST, and NSFC (China); COLCIENCIAS (Colombia); MSES (Croatia); RPF 
(Cyprus); Academy of Sciences and NICPB (Estonia); Academy of Finland, 
ME, and HIP (Finland); CEA and CNRS/IN2P3 (France); BMBF, DFG, and HGF 
(Germany); GSRT (Greece); OTKA and NKTH (Hungary); DAE and DST (India); 
IPM (Iran); SFI (Ireland); INFN (Italy); NRF and WCU (Korea); LAS 
(Lithuania); CINVESTAV, CONACYT, SEP, and UASLP-FAI (Mexico); PAEC 
(Pakistan); SCSR (Poland); FCT (Portugal); JINR (Armenia, Belarus, 
Georgia, Ukraine, Uzbekistan); MST and MAE (Russia); MSTD (Serbia); 
MICINN and CPAN (Spain); Swiss Funding Agencies (Switzerland); NSC 
(Taipei); TUBITAK and TAEK (Turkey); STFC (United Kingdom); DOE and NSF 
(USA). 

Individuals have received support from the Marie-Curie programme and the 
European Research Council (European Union); the Leventis Foundation; the 
A. P. Sloan Foundation; the Alexander von Humboldt Foundation; the 
Associazione per lo Sviluppo Scientifico e Tecnologico del Piemonte 
(Italy); the Belgian Federal Science Policy Office; the Fonds pour la 
Formation \`a la Recherche dans l'Industrie et dans l'Agriculture 
(FRIA-Belgium); and the Agentschap voor Innovatie door Wetenschap en 
Technologie (IWT-Belgium).

\bibliography{auto_generated}   
\cleardoublepage \appendix\section{The CMS Collaboration \label{app:collab}}\begin{sloppypar}\hyphenpenalty=5000\widowpenalty=500\clubpenalty=5000\textbf{Yerevan Physics Institute,  Yerevan,  Armenia}\\*[0pt]
S.~Chatrchyan, V.~Khachatryan, A.M.~Sirunyan, A.~Tumasyan
\vskip\cmsinstskip
\textbf{Institut f\"{u}r Hochenergiephysik der OeAW,  Wien,  Austria}\\*[0pt]
W.~Adam, T.~Bergauer, M.~Dragicevic, J.~Er\"{o}, C.~Fabjan, M.~Friedl, R.~Fr\"{u}hwirth, V.M.~Ghete, J.~Hammer\cmsAuthorMark{1}, S.~H\"{a}nsel, M.~Hoch, N.~H\"{o}rmann, J.~Hrubec, M.~Jeitler, W.~Kiesenhofer, M.~Krammer, D.~Liko, I.~Mikulec, M.~Pernicka, B.~Rahbaran, H.~Rohringer, R.~Sch\"{o}fbeck, J.~Strauss, A.~Taurok, F.~Teischinger, C.~Trauner, P.~Wagner, W.~Waltenberger, G.~Walzel, E.~Widl, C.-E.~Wulz
% ...
\vskip\cmsinstskip
% ...
1:~~Also at CERN, European Organization for Nuclear Research, Geneva, Switzerland\\
% ...

\end{sloppypar}
\end{document}